# Understanding and Shaping Human-Technology Assemblages in the Age of Generative AI

Josh Andres
The Australian National University
(first.last@anu.edu.au)

Chris Danta
The Australian National University
(first.last@anu.edu.au)

Andrea Bianchi
The Korea Advanced Institute of Science and Technology KAIST
(andrea@kaist.id)

Sungyeon Hong
The Australian National University
(first.last@anu.edu.au)

Zhuying Li
Southeast University, China
(zhuyingli@seu.edu.cn)

Eduardo B. Sandoval
University of New South Wales, Australia
(e.sandoval@unsw.edu.au)

Charles Martin
The Australian National University
(first.last@anu.edu.au)

Ned Cooper
The Australian National University
(first.last@anu.edu.au)

## ABSTRACT

Generative AI capabilities are rapidly transforming how we perceive, interact with, and relate to machines. This one-day workshop invites HCI researchers, designers, and practitioners to imaginatively inhabit and explore the possible futures that might emerge from humans combining generative AI capabilities into everyday technologies at massive scale. Workshop participants will craft stories, visualisations, and prototypes through scenario-based design to investigate these possible futures, resulting in the production of an open-annotated scenario library and a journal or interactions article to disseminate the findings. We aim to gather the DIS community knowledge to explore, understand and shape the relations this new interaction paradigm is forging between humans, their technologies and the environment in safe, sustainable, enriching, and responsible ways.

**ACM Reference Format:**

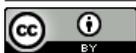

## 1 BACKGROUND AND MOTIVATION

The ability of Generative AI to engage in conversations, create media, and orchestrate tasks across various software and devices signifies an increased level of agency for machines. This transformation directly impacts how we perceive, engage, and connect with machines, shaping the dynamics and the nature of human-technology relationships. Specific examples include novel models like ChatGPT which can see, hear, and converse, providing follow-up questions, redirecting discussions, connecting related concepts, changing arguments and topics, and assuming different roles and situations [17]. As capabilities like these are combined with everyday technologies, including personal devices, home appliances, robots, classrooms, homes, vehicles, and entertainment technologies, and they propagate across society, they can influence our perceptions of certain topics and shape creative exploration and critical thinking. It is vital today to recognize these evolving dynamics to explore their future through design.

As billions of people across different generations and geographies interact everyday with generative AI in physical devices and artifacts, humans and machines begin to coevolve in new ways. This one-day workshop explores the possible futures that might develop from humans combining generative AI capabilities with everyday technologies at massive scale. It investigates the emergent assemblages forming between humans and technologies imbued with generative capabilities, and the roles played by, attributed to, implicit and explicit in these assemblages [8]. Our starting assumption is that the new interaction paradigm will transform relationships between humans, their technologies, society, and the environment. Understanding and shaping human-machine coevolution through design enables us to support human and nonhuman agencies, values-oriented design, social change and care for the environment.

We believe that the combination of generative AI capabilities into everyday technologies calls for new ways of thinking about the human-technology relations. Paradigms such as interaction, where human intent predominantly dictates machine responses, are characterised by a directional influence from human to machine [9]. This is also the case for augmentation technologies that extend human cognitive, sensory, and physical abilities [1, 2, 4, 14]. Wearable augmentation, such as exoskeletons and prostheses transition from being merely tools to becoming extensions of the human body [13]. Meanwhile, nonwearable augmentation technologies like mobile phones, GPS, and calculators are often used transactionally to augment cognition. In both interaction and augmentation, the machine's operation is directed by the user's input without any additional interpretation or decision-making by the machine itself.

Interaction and augmentation suggest that our relationships with machines mainly involve a unidirectional influence "from humans to machines," but this dynamic is changing to become more bidirectional "between humans and machines," with both actively influencing one another. For example, humans influence machines through the objectives, data, rules, and algorithmic behaviours that ultimately shape their actions. On a daily basis, machines gently influence human thinking and decision-making through recommendation systems, social media feeds, and voice assistants, reshaping habits and choices. In relations involving humans and technologies imbued with generative AI capabilities, a bidirectional influence is often at play. As humans provide more input data and feedback, their behaviour changes. Equally, as technology evolves, it offers novel directions and uses from humans - creating a feedback loop of mutual influence. When this bidirectional influence unfolds, we witness the becoming of the assemblage—a human-technology relationship that engages with different temporalities, ranging from experiential to computationally measurable, and embodies multiple ways of sensing and knowing across various spatialities. This necessitates an emphasis on the roles played by both the human and the technology, as well as the assemblage itself, within a given context. As assemblages evolve, they present the potential for further evolution in our relationships with technology.

## 2 EXTENDING SCENARIO-BASED DESIGN TO ATTEND TO RELATIONSHIP DYNAMICS

To explore the assemblages forming between humans and technologies imbued with generative capabilities, we propose an extension of scenario-based design methodologies. Scenario-based design envisions and interrogates possible futures shaped by technology [6, 11, 12, 15], often materialised through scenarios, stories, visualisations, or prototypes, embodying design thinking, research insights, and speculative exploration in a form that can be shared, discussed, and built upon. In HCI, scenario-based design has been used to envision the use and implications of emerging technologies. Works by Carroll [5], established a framework for using narratives early in the design and development processes. Others explored the ethical, social, and environmental implications of technology use, as seen in the speculative design works of Dunne & Raby [7]. More recent examples used scenario-based design to explore and inform the design of explainable AI (XAI) technologies, ranging from humanitarian aid [3] to supporting coding [16].

Scenario-based design draws on various ingredients to depict evocative illustrations of the future [5, 10]. Common ingredients include *a setting or situation state*, such as a place and policy regulation; *one or more actors*, including humans, nonhumans (animals, lakes, spirits, etc.) and machines; *agendas* that supply the underlying motivations for actors to act; and *props*, including tools and objects. Often, scenarios describe a sequence of interactions or inspire artifacts as forms of enquiry. Other ingredients used to explore systemic aspects of scenarios are *pervasiveness*, focusing on the *what-if* of technological adoption; *systemic effects*, referring to environmental, social, and cultural impacts of technology; and *temporality*, exploring what the world would be like five, ten, or twenty years after a technology has been deployed.

We propose to introduce another ingredient to scenario-based design, namely, "*relationship dynamics,*" to explore the roles played by, attributed to, implicit, and explicit in assemblages between humans and technologies. This new vital ingredient focuses attention on the types of relationships facilitated, their characteristics, and their evolution.

Our goal is to gather and share knowledge from the DIS community and beyond to develop, share, reflect on, and synthesize insights from scenarios and discussions. These insights will focus on the design of artifacts and interventions aimed at purposefully shaping our futures, as we consider combining (or not) generative AI capabilities with everyday technologies.

## 3 WORKSHOP THEMES

We aim to explore the below themes and questions in this workshop:

**Evolving relations between people and AI** - the profound impact of generative AI capabilities presents the potential to redefine relations between humans, machines and the environment. Given the rapid addition of these capabilities into everyday life, it is timely to explore the role of design in creating possibilities to explore these relations.

- What explicit or implicit roles should generative AI-enabled technologies play to contribute positively to the human-technology relationship?
- How can we design for reciprocal influence between humans and machines to foster human growth and technological advancements without leading to over-dependency?
- In the human-technology relationship, how shall our technologies age with us?

**Interactional to systemic perspectives** - scenarios can vividly captures depictions ranging from interactional to pervasive scales, illustrating technologies' short and long-term effects in a sociocultural, ethical, and environmental context.

- How can we design generative AI-enabled technologies that reveal their ongoing societal and environmental impacts over time and encourage reflection/action?
- Considering the societal and environmental trajectories, should generative AI-enabled technologies prepare us for, or guide us towards particular futures?

These themes and questions can support our community in engaging with the multifaceted impact of designing technologies that incorporate generative AI capabilities, considering interactional, systemic, temporal, and relational dimensions.

## 4 WORKSHOP GOALS AND ANTICIPATED OUTCOMES

We aim to foster and curate a global, diverse forum of designers, researchers, and practitioners that ranges in career stage and epistemological perspectives to explore the dynamic assemblages between humans and generative AI-enabled technologies. We will investigate how this exciting paradigm can offer new relations to ourselves, our technologies, societies, and the environment.

The primary objectives to support the research efforts are: 1) Foster and curate a multidisciplinary network for ongoing collaboration to study and shape this paradigm by learning from various applications, including design research approaches. 2) Share and reflect on design practices based on participants' engagements with specific technologies, situations, and people to deepen our understanding of how and how not to combine and imbue everyday technologies with generative AI capabilities. 3) Generate new knowledge as insights, practical strategies, and recommendations about the novel human-technology assemblages and their dynamics facilitated by this paradigm. 4) Create an annotated open library to share with the community the scenario-based designs that participants will bring to the workshop and those created during the workshop. To capture tacit design research knowledge and the reflections that emerge throughout the day. The workshop's scenarios, reflections, and synthesis will be disseminated through the annotated open library to extend beyond an academic platform and engage a broader audience, as well as through a journal or interactions article.

## 5 WORKSHOP ACTIVITIES

- Applicants are invited to craft a scenario-based design to participate in the workshop, exploring one or multiple of the themes and employing a combination of scenario ingredients, with "*relationship dynamics*" as a compulsory ingredient. Participants are encouraged to draw on technologies they research and/or use regularly.
- During the workshop, participants will show and tell their scenarios, which can be stories, visualisations, and prototypes - we will collaboratively collect notes and reflections to begin the scenario annotation process.
- After the scenario showcase, a panel specialising in the evolution of human-technology relations will share their insights. This will serve to both contrast with and expand the scenarios discussed.
- In groups, participants will collaboratively create new scenarios, informed by the collective reflections about the showcase and panel discussions. This process will offer participants opportunities for design making through various mediums, including written text, arts and crafts, props, and components of sensing technology.
- Participants will share their scenarios, reflecting and synthesizing ideas and themes as future design research on the human-technology assemblages in safe, sustainable, enriching, and responsible ways.
- During the final two hours, we will utilise the insights gathered from the morning and early afternoon sessions in a guided tour through the Design Museum Denmark, examining the scenarios and relationships hinted at by the exhibited artifacts. This activity will offer an opportunity to apply and extend the workshop's insights. The day will conclude with a reflection session on these experiences over dinner, enriching our understanding of designing for this new paradigm's potential in everyday life.

## ACKNOWLEDGMENTS

Andrea Bianchi was supported by the National Research Foundation of Korea (NRF) grant funded by the Korean government (MSIT) No. 2018R1A5A7025409. Zhuying Li was supported by the National Natural Science Foundation of China No. 62302094. Chris Danta was supported by the Australian Research Council Future Fellowship FT200100914.

## REFERENCES

[1] Josh Andres, Tuomas Kari, Juerg Von Kaenel, and Florian'Floyd' Mueller. 2019. Co-riding With My eBike to Get Green Lights. In *Proceedings of the 2019 on Designing Interactive Systems Conference*, 2019. 1251–1263.
[2] Josh Andres, m.c. schraefel, Nathan Semertzidis, Brahmi Dwivedi, Yutika C. Kulwe, Juerg von Kaenel, and Florian Floyd Mueller. 2020. Introducing Peripheral Awareness as a Neurological State for Human-Computer Integration. In *Proceedings of the 2020 CHI Conference on Human Factors in Computing Systems* (*CHI '20*), 2020. Association for Computing Machinery, New York, NY, USA, 1–13. https://doi.org/10.1145/3313831.3376128
[3] Josh Andres, Christine T. Wolf, Sergio Cabrero Barros, Erick Oduor, Rahul Nair, Alexander Kjærum, Anders Bech Tharsgaard, and Bo Schwartz Madsen. 2020. Scenario-based XAI for Humanitarian Aid Forecasting. In *Extended Abstracts of the 2020 CHI Conference on Human Factors in Computing Systems* (*CHI EA '20*), April 25, 2020. Association for Computing Machinery, New York, NY, USA, 1–8. https://doi.org/10.1145/3334480.3382903
[4] Zahra Ashktorab, Michael Desmond, Josh Andres, Michael Muller, Narendra Nath Joshi, Michelle Brachman, Aabhas Sharma, Kristina Brimijoin, Qian Pan, Christine T. Wolf, Evelyn Duesterwald, Casey Dugan, Werner Geyer, and Darrell Reimer. 2021. AI-Assisted Human Labeling: Batching for Efficiency without Overreliance. *Proc. ACM Hum.-Comput. Interact.* 5, CSCW1 (April 2021), 89:1-89:27. https://doi.org/10.1145/3449163
[5] John M. Carroll. 2003. *Making Use: Scenario-Based Design of Human-Computer Interactions*. MIT Press.
[6] Jialin Deng, Patrick Olivier, Josh Andres, Kirsten Ellis, Ryan Wee, and Florian Floyd Mueller. 2022. Logic Bonbon: Exploring Food as Computational Artifact. In *Proceedings of the 2022 CHI Conference on Human Factors in Computing Systems* (*CHI '22*), April 28, 2022. Association for Computing Machinery, New York, NY, USA, 1–21. https://doi.org/10.1145/3491102.3501926
[7] Anthony Dunne and Fiona Raby. 2013. *Speculative Everything: Design, Fiction, and Social Dreaming*. MIT Press.
[8] N. Katherine Hayles. 2017. 5. Cognitive Assemblages: Technical Agency and Human Interactions. In *5. Cognitive Assemblages: Technical Agency and Human Interactions*. University of Chicago Press, 115–141. https://doi.org/10.7208/9780226447919-007
[9] Kasper Hornbæk and Antti Oulasvirta. 2017. What Is Interaction? In *Proceedings of the 2017 CHI Conference on Human Factors in Computing Systems* (*CHI '17*), May 02, 2017. Association for Computing Machinery, New York, NY, USA, 5040–5052. https://doi.org/10.1145/3025453.3025765
[10] Lisa P. Nathan, Predrag V. Klasnja, and Batya Friedman. 2007. Value scenarios: a technique for envisioning systemic effects of new technologies. In *CHI '07 Extended Abstracts on Human Factors in Computing Systems* (*CHI EA '07*), April 28, 2007. Association for Computing Machinery, New York, NY, USA, 2585–2590. https://doi.org/10.1145/1240866.1241046
[11] William Odom, MinYoung Yoo, Henry Lin, Tijs Duel, Tal Amram, and Amy Yo Sue Chen. 2020. Exploring the Reflective Potentialities of Personal Data with Different Temporal Modalities: A Field Study of Olo Radio. In *Proceedings of the 2020 ACM Designing Interactive Systems Conference* (*DIS '20*), July 03, 2020. Association for Computing Machinery, New York, NY, USA, 283–295. https://doi.org/10.1145/3357236.3395438
[12] Doros Polydorou, Kening Zhu, and Antje Illner. "This is a techno-necklace from my great grandmother": Animism-Inspired Design Guidelines for Digitally Ensouled Jewellery. *Hong Kong*.

[13] Roope Raisamo, Ismo Rakkolainen, Päivi Majaranta, Katri Salminen, Jussi Rantala, and Ahmed Farooq. 2019. Human augmentation: Past, present and future. *International Journal of Human-Computer Studies* 131, (2019), 131–143.

[14] Nathan Semertzidis, Michaela Scary, Josh Andres, Brahmi Dwivedi, Yutika Chandrashekhar Kulwe, Fabio Zambetta, and Florian Floyd Mueller. 2020. Neo-Noumena: Augmenting Emotion Communication. In *Proceedings of the 2020 CHI Conference on Human Factors in Computing Systems*. Association for Computing Machinery, New York, NY, USA, 1–13. Retrieved March 24, 2022 from https://doi.org/10.1145/3313831.3376599

[15] Irina Shklovski and Erik Grönvall. 2020. CreepyLeaks: Participatory Speculation Through Demos. In *Proceedings of the 11th Nordic Conference on Human-Computer Interaction: Shaping Experiences, Shaping Society* (*NordiCHI '20*), October 26, 2020. Association for Computing Machinery, New York, NY, USA, 1–12. https://doi.org/10.1145/3419249.3420168

[16] Jiao Sun, Q. Vera Liao, Michael Muller, Mayank Agarwal, Stephanie Houde, Kartik Talamadupula, and Justin D. Weisz. 2022. Investigating Explainability of Generative AI for Code through Scenario-based Design. In *27th International Conference on Intelligent User Interfaces* (*IUI '22*), March 22, 2022. Association for Computing Machinery, New York, NY, USA, 212–228. https://doi.org/10.1145/3490099.3511119

[17] Tianyu Wu, Shizhu He, Jingping Liu, Siqi Sun, Kang Liu, Qing-Long Han, and Yang Tang. 2023. A brief overview of ChatGPT: The history, status quo and potential future development. *IEEE/CAA Journal of Automatica Sinica* 10, 5 (2023). https://ieeexplore.ieee.org/abstract/document/10113601